\documentstyle[prl,aps,multicol]{revtex}
\begin{document}
\draft

\title{Renormalization Group Theory for a Perturbed KdV Equation}

\author{Tao Tu,$^{1}\thanks{
Corresponding author will be addressed: tutao@mail.ustc.edu.cn}$ and Hua
Sheng$^{2}$}
\address{$^{1}$Center of Nonlinear Science,\\
University of Science and Technology of China, Hefei, Anhui, 230026,
P.R.China\\
$^{2}$Department of Physics,\\
University of Maryland, College park, MD 20742 U.S.A \\
}
\date{6 Jun 2001}

\maketitle

\begin{abstract}
We show that renormalization group(RG) theory can be used to give an
analytic description of the evolution of a perturbed KdV equation. The
equations describing the deformation of its shape as the effect of
perturbation are RG equations. The RG approach may be simpler than inverse
scattering theory(IST) and another approaches, because it dose not rely on
any knowledge of IST and it is very concise and easy to understand. To the
best of our knowledge, this is the first time that RG has been used in this
way for the perturbed soliton dynamics.
\end{abstract}

%%%%%%%%%%%%%%%%%%%%%%%%%%%%%%%%%main body%%%%%%%%%%%%%%%%%%%%%%%%%%%%
\begin{multicols}{2}

The standard soliton equations(KdV, NLS, Sine-Gordon and so on.)are highly
idealized. In more realistic situations, it is important to understand
nonlinear evolution equations under the influence of perturbations.$%
^{\lbrack 1-2\rbrack }$ \ Several different approaches to the analytic
description of soliton dynamics in these perturbed systems are known. The
most powerful method to deal with these cases is based on IST.$^{\lbrack
3-4\rbrack }$ It turns out that for any nonlinear evolution equation which
is reasonably close to a nonlinear evolution equation that can be exactly
solved by IST, the total evolution of the scattering data can be given and
also be expanded in a perturbation expansion. Thus one can determine the
effects of perturbations such as small dissipations, relaxations, etc., on
the evolution of these exactly solvable nonlinear evolution equations, and
in particular, the effects on the soliton states. These equations which
describe the time dependence of the scattering data can be solved in the
''adiabatic'' approximation. But it is inconvenient \ to use for one who is
not familiar with IST. Another alternative way to study soliton
perturbations is so-called direct perturbation theory based on a Green's
function and a rather standard 'two-timing' procedure.$^{\lbrack 5-8\rbrack }
$ In this scheme, a basic technical ingredient is to construct a Green's
function. Then the time dependence of the soliton parameters and the
first-order correction are readily available. But they stated that their
method partially requires the knowledge of IST.

In this paper we perform the RG approach to the perturbed KdV equation and
compare the result with another methods. We demonstrate that an analytic
description of the evolution of a soliton, including the deformation of its
shape and tail formation can be given, and the equations describing the
evolution of the soliton amplitude and velocity as the effect of
perturbation are RG equations.The apparent advantage of the RG approach is
that the starting point is simply a straightforward naive perturbation
expansion, for which very little a $priori$ knowledge is required. We also
see that the RG approach may be more efficient and concise in practice than
another methods.

The KdV equation

\begin{equation}
\vspace{1pt}u_{t}+6uu_{x}+u_{xxx}=0
\end{equation}
has a single soliton solution such as

\begin{equation}
u=2A_{0}^{2}\sec h^{2}z,\text{ \ \ }z=A_{0}(x-\xi ),\text{ \ \ }\xi
=4A_{0}^{2}t+\xi _{0}
\end{equation}
where $A_{0}$ is the amplitude and $\xi _{0}$ is the initial position of the
soliton.

Now we consider the perturbed KdV equation$^{\lbrack 9\rbrack }$ 
\begin{equation}
\vspace{1pt}u_{t}+6uu_{x}+u_{xxx}=\varepsilon R\lbrack u\rbrack 
\end{equation}
with the initial condition 
\begin{equation}
u(x,0)=2A_{0}^{2}\sec h^{2}A_{0}(x-\xi _{0})
\end{equation}
To proceed, we construct a $\varepsilon -$perturbation to this equation 
\begin{equation}
u=u_{0}+\varepsilon u_{1}+....
\end{equation}
Put it into Eq.(3), we have 
\begin{equation}
\vspace{1pt}u_{0t}+6u_{0}u_{0x}+u_{0xxx}=0
\end{equation}

\begin{equation}
u_{1t}+6u_{0}u_{1x}+6u_{0x}u_{1}+u_{1xxx}=R\lbrack u_{0}\rbrack
\end{equation}
\begin{equation}
u_{2t}+6u_{0}u_{2x}+6u_{0x}u_{2}+u_{2xxx}=R\lbrack u_{0},u_{1}\rbrack
-6u_{1}u_{1x}
\end{equation}
and so on. The initial condition now changes into 
\begin{equation}
u_{0}(x,0)=2A_{0}^{2}\sec h^{2}A_{0}(x-x_{0})
\end{equation}
\begin{equation}
u_{n}(x,0)=0,\text{ \ \ }n\geq 1
\end{equation}
In the first-order, we write 
\begin{equation}
\lbrack \partial _{t}-\stackrel{\symbol{94}}{L}\rbrack u_{1}=R\lbrack
u_{0}\rbrack
\end{equation}
where 
\begin{equation}
\stackrel{\symbol{94}}{L}=6u_{0}\partial _{x}+6u_{0x}+\partial _{xxx}.
\end{equation}
For convenience, we use $z$ instead of $x$. Then Eq.(12) becomes 
\begin{equation}
\stackrel{\symbol{94}}{L}=A_{0}^{3}\lbrack \frac{d^{3}}{dz^{3}}+(12\sec
h^{2}z-4)\frac{d}{dz}-24\tanh z\sec h^{2}z\rbrack
\end{equation}

We can derive the solution to this linear partial differential equation (11)
by the method of variables separation:$^{\lbrack 10-12\rbrack }$ 
\begin{equation}
u_{1}(z,t)=u_{1}^{s}+regular\text{ }terms
\end{equation}
so that the singular part of $u_{1}$ is given by 
\begin{equation}
u_{1}^{s}=f_{1}t\Phi _{1}(z)+(4A_{0}^{3}f_{1}t^{2}+f_{2}t)\Phi _{2}(z)
\end{equation}
\begin{equation}
f_{1}=\int_{-\infty }^{+\infty }R\lbrack u_{0}\rbrack \Psi _{1}(z,k)dz
\end{equation}
\begin{equation}
f_{2}=\int_{-\infty }^{+\infty }R\lbrack u_{0}\rbrack \Psi _{2}(z,k)dz
\end{equation}
\begin{equation}
regular\text{ }terms=\int_{-\infty }^{+\infty }G(z,y;t)R\lbrack u_{0}\rbrack
dy
\end{equation}
\begin{equation}
G(z,y;t)=-P\int_{-\infty }^{+\infty }\frac{1-\exp \lbrack
ik(k^{2}+4)A_{0}^{3}t\rbrack }{ik(k^{2}+4)A_{0}^{3}}\Phi (z,k)\Psi (y,k)dk
\end{equation}
where 
\begin{eqnarray}
\Phi (z,k) &=&\frac{1}{\sqrt{2\pi }k(k^{2}+4)}\lbrack
k(k^{2}+4)+4i(k^{2}+2)\tanh z  \nonumber \\
&&-8k\tanh ^{2}z-8i\tanh ^{3}z\rbrack \exp \lbrack ikz\rbrack
\end{eqnarray}

\begin{equation}
\Phi _{1}(z,k)=(1-z\tanh z)\sec h^{2}z
\end{equation}

\begin{equation}
\Phi _{2}(z,k)=\tanh z\sec h^{2}z
\end{equation}
\begin{equation}
\Psi (z,k)=\frac{1}{\sqrt{2\pi }k(k^{2}+4)}\lbrack k^{2}-4ik\tanh z-4\tanh
^{2}z\rbrack \exp \lbrack ikz\rbrack 
\end{equation}
\begin{equation}
\Psi _{1}(z,k)=\sec h^{2}z
\end{equation}
\begin{equation}
\Psi _{2}(z,k)=\tanh z+z\sec h^{2}z
\end{equation}

Thus, the bare perturbation result is 
\begin{eqnarray}
u(z,t) &=&2A_{0}^{2}\sec h^{2}z+\varepsilon \lbrack f_{1}t\Phi
_{1}(z)+(4A_{0}^{3}f_{1}t^{2}+f_{2}t)\Phi _{2}(z)\rbrack   \nonumber \\
&&+regular\text{ }terms\text{ \ \ }  \nonumber \\
z &=&A_{0}(x-\xi ),\text{ \ \ }\xi =4A_{0}^{2}t+\xi _{0}
\end{eqnarray}
which is divergent when $t\rightarrow \infty $ as anticipated. To deal with
the divergence, we use the RG(renormalization group) method which has been
introduced by Goldenfeld et. al for global asymptotic analysis.$^{\lbrack
13-14\rbrack }$ First, to regularize the naive perturbation series, we
introduce an arbitrary time $\tau $ and split $t$ as $t-\tau +\tau $. The
quantity $A_{0}$ and $\xi _{0},$ the ''initial amplitude'' and ''initial
position'' of the soliton, can not be obtained from the knowledge of $u(z,t)$
at large times, therefore $A_{0}$ and $\xi _{0}$ are considered to be
''unobservable'' at large times in the same way that the ''bare'' electric
charge is unobservable at long distances according to quantum
electrodynamics.$^{\lbrack 15\rbrack }$ In another words, $A_{0}$ and $\xi
_{0}$ are no longer constants of motion in the presence of the perturbation.
We cure this divergence of the bare perturbation series by introducing the
renormalized amplitude and position 
\begin{equation}
A_{R}=Z_{1}^{-1}(\tau )A_{0}
\end{equation}
\begin{equation}
\xi _{R}=Z_{2}^{-1}(\tau )\xi _{0}
\end{equation}
The multiplicative renormalization constants $Z_{1}$ and $Z_{2}$ are
introduced to absorb the terms $\tau $. We proceed by assuming the expansion 
\begin{equation}
Z_{1}=1+\sum_{1}^{\infty }a_{n}\varepsilon ^{n}
\end{equation}
\begin{equation}
Z_{2}=1+\sum_{1}^{\infty }b_{n}\varepsilon ^{n}
\end{equation}
The coefficients $a_{n}$ and $b_{n}$ are chosen to eliminate the terms
containing $\tau $ order by order in $\varepsilon $. In the first order, we
obtain 
\begin{equation}
a_{1}=-\frac{f_{1}\tau }{4A_{R}^{2}}
\end{equation}
\begin{equation}
b_{1}=-\frac{f_{2}\tau }{4A_{R}^{3}x_{R}}
\end{equation}
and we have the following renormalization perturbation result 
\begin{eqnarray}
u_{R} &=&2A_{R}^{2}\sec h^{2}z+\varepsilon \{f_{1}(t-\tau )\Phi _{1}(z)+ 
\nonumber \\
&&\lbrack 4A_{R}^{3}f_{1}(t^{2}-\tau ^{2})+f_{2}(t-\tau )\rbrack \Phi
_{2}(z)\}+  \nonumber \\
&&regular\text{ }terms.
\end{eqnarray}
where $A_{R}$ and $\xi _{R}$\ are now functions of $\tau $. Since $\tau $
does not appear in the original problem, the solution should not depend on
it. That is to say $\left( \frac{\partial u_{R}}{\partial \tau }\right)
_{t}=0$ which is the RG argument of Gell-Mann and Low.$^{\lbrack 16\rbrack }$
We get the following RG equation 
\begin{equation}
\frac{dA_{R}}{d\tau }=\varepsilon \frac{1}{4A_{R}}f_{1}=\varepsilon \frac{1}{%
4A_{R}}\int_{-\infty }^{+\infty }R\lbrack u_{0}\rbrack \sec h^{2}zdz
\end{equation}
\begin{equation}
\frac{d\xi _{R}}{d\tau }=\varepsilon \frac{1}{4A_{R}^{3}}f_{2}=\varepsilon 
\frac{1}{4A_{R}^{3}}\int_{-\infty }^{+\infty }R\lbrack u_{0}\rbrack (\tanh
z+z\sec h^{2}z)dz
\end{equation}
which in this case consists of two independent equations. This is analogous
to the RG equation in field theory. Finally, setting $\tau =t$ in Eq.(33)
eliminates the secular term, giving the result

\begin{eqnarray}
u(x,t) &=&2A^{2}\sec h^{2}z+regular\text{ }terms,\text{\ \ }  \nonumber \\
A &=&A_{R},\text{ \ \ }z=A(x-\xi ),\text{ \ \ }\xi =4A^{2}t+\xi _{R},
\end{eqnarray}
where the regular terms are the 'tail'' formation$^{\lbrack 3-4\rbrack }$
and 
\begin{equation}
\frac{dA}{dt}=\varepsilon \frac{1}{4A}\int_{-\infty }^{+\infty }R\lbrack
u_{0}\rbrack \sec h^{2}zdz
\end{equation}
\begin{equation}
\frac{d\xi }{dt}=4A^{2}+\varepsilon \frac{1}{4A^{3}}\int_{-\infty }^{+\infty
}R\lbrack u_{0}\rbrack (\tanh z+z\sec h^{2}z)dz
\end{equation}
The two important equations which determine how the soliton shape and
position are affected by the perturbation are also consistent with those
derived by IST.$^{\lbrack 3-4\rbrack }$

As an example, we consider the damping KdV equation in which $R\lbrack
u\rbrack =-u$. The time dependence of the soliton parameters can be easily
obtained from Eq.(37) and Eq.(38). Thus we have 
\begin{equation}
A=A_{0}\exp (-\frac{2\varepsilon t}{3})
\end{equation}
\begin{equation}
\xi =\frac{3}{\varepsilon }A_{0}^{2}\lbrack 1-\exp (-\frac{4\varepsilon t}{3}%
)\rbrack
\end{equation}
which is just the same as that obtained by IST.$^{\lbrack 17\rbrack }$

Then we consider another example, the KdV-Burgers equation 
\begin{equation}
u_{t}+6uu_{x}+u_{xxx}=\varepsilon u_{xx}
\end{equation}
In the same way, we obtain 
\begin{equation}
A=\frac{A_{0}}{\sqrt{1+\frac{16}{15}\varepsilon tA_{0}^{2}}}
\end{equation}
\begin{equation}
\xi =\frac{15}{4\varepsilon }\log (1+\frac{16}{15}\varepsilon tA_{0}^{2})
\end{equation}
which is also the same as that derived by IST.$^{\lbrack 18\rbrack }$

In summary, we have demonstrated that a perturbed KdV equation can be solved
by RG\ method, with some attendant technical advantages compared with other
methods. The present approach can easily be generalized also to
multiple-soliton state. To avoid unnecessary complications we expound our
theory using as example the KdV equation. It is, however, clear from what
follows that this does not restrict the general nature of the method for
another perturbed evolution equations.

Tao Tu is supported in part by the National Science Foundation in China
(No.19875047). The authors would like to thank Dr. Liu Jian Wei, Professor
G. Cheng, Professor Huang. N. N for their help in this paper. One of \ the
authors (Tao Tu) gratefully thanks Professor N. Goldenfeld for consideration of his work. The authors also wishes to thank two anonymous
referees for their criticism and suggestions.

\end{multicols}{2}


\begin{references}
\bibitem{}  There also exists another approaches which would be mentioned
here: the modified conservation laws, the Hamiltonian and Lagrangian
formalism, and a canonical transformations, etc. But they are of limited use
than IST.
\end{references}
\end{document}